\documentclass[sigconf]{acmart}
\AtBeginDocument{%
  }

\copyrightyear{2026}
\acmYear{2026}
\setcopyright{cc}
\setcctype{by}
\acmConference[CHI EA '26]{Extended Abstracts of the 2026 CHI Conference on Human Factors in Computing Systems}{April 13--17, 2026}{Barcelona, Spain}
\acmBooktitle{Extended Abstracts of the 2026 CHI Conference on Human Factors in Computing Systems (CHI EA '26), April 13--17, 2026, Barcelona, Spain}
\acmDOI{10.1145/3772363.3799190}
\acmISBN{979-8-4007-2281-3/2026/04}

\begin{document}

\title{A Resource-Rational Principle for Modeling Visual Attention Control}


\author{Yunpeng Bai}
\affiliation{%
  \institution{National University of Singapore}
  \country{Singapore}}

\renewcommand{\shortauthors}{Yunpeng Bai}

\begin{abstract}
  Understanding how people allocate visual attention is central to Human-Computer Interaction (HCI), yet existing computational models of attention are often either descriptive, task‑specific, or difficult to interpret. My dissertation develops a resource‑rational, simulation‑based framework for modeling visual attention as a sequential decision‑making process under perceptual, memory, and time constraints. I formalize visual tasks, such as reading and multitasking, as bounded‑optimal control problems using Partially Observable Markov Decision Processes, enabling eye‑movement behaviors, such as fixation and attention switching to emerge from rational adaptation rather than being hand‑coded or purely data‑driven. These models are instantiated in simulation environments spanning traditional text reading and reading‑while‑walking with smart glasses, where they reproduce classic empirical effects, explain observed trade‑offs between comprehension and safety, and generate novel predictions under time pressure and interface variation. Collectively, this work contributes a unified computational account of visual attention, offering new tools for theory‑driven and resource‑efficient HCI design.
\end{abstract}




\maketitle

\section{Introduction}

Visual attention is a primary channel for acquiring information~\cite{posner1980orienting,desimone1995neural,carrasco2011visual}. It shapes how people perceive and act in everyday environments, including interactive systems~\cite{land2001ways,hayhoe2005eye}. Beyond sampling visual input, attention allocation structures how users coordinate perception with action during device interaction~\cite{shi2024crtypist,shi2025chartist,shi2025simulating,jokinen2021touchscreen} and supports contextual awareness and safety in multitasking~\cite{bai2024heads,lingler2024supporting}. Despite its centrality, there is still no systematic computational account of how humans control visual attention across tasks and interaction contexts~\cite{bai2024heads}.

Simulation offers a principled route toward such an account by enabling mechanistic hypotheses about attention control to be formalized and tested~\cite{murray2022simulation,oulasvirta2022computational,bairesource,bai2024heads}. However, classic models often rely on handcrafted policies (e.g., E-Z Reader, SWIFT)~\cite{reichle2003ez,engbert2005swift}, which limits their adaptability to diverse tasks, environments, and interface designs~\cite{oulasvirta2022computational,bairesource}. Recent data-driven approaches achieve strong predictive performance (e.g., ScanDL, EyeTtention, VQA-based models)~\cite{bolliger2023scandl,Bolliger2025ScanDL2,deng2023eyettention,chen2021predicting,jiang2023ueyes,jiang2024eyeformer,jiang2024graph4gui}, but typically require large-scale training data and prioritize behavioral imitation over explaining \emph{why} attention is allocated as observed~\cite{bairesource}.

My dissertation addresses these limitations by modeling visual attention as \emph{resource-rational control}. I formalize users as boundedly rational agents who select attentional actions to maximize task utility while trading off cognitive and environmental costs under limited perceptual and cognitive resources~\cite{lieder2020resource,howes2009rational,oulasvirta2022computational}. Rather than prescribing fixed heuristics, this principle allows attentional strategies to emerge as bounded-optimal policies adapted to task demands and constraints. Concretely, I formulate attention control as sequential decision-making under uncertainty using partially observable Markov decision processes (POMDPs)~\cite{sutton1998reinforcement}, which provide an explicit and interpretable structure in terms of states, observations, transitions, and rewards. Policies are learned using deep reinforcement learning (PPO)~\cite{schulman2017proximal}, enabling agents to acquire attention strategies through interaction with simulated environments rather than supervision from large-scale human gaze data. While the core modeling framework is established, key aspects of the dissertation, particularly the framing of simulation-based modeling as a design methodology for HCI, and the treatment of individual and group differences remain open and are the focus of ongoing refinement.

\section{Research Objectives and Methodology}

\begin{figure*}[t]
  \centering 
  \includegraphics[width=1\textwidth]{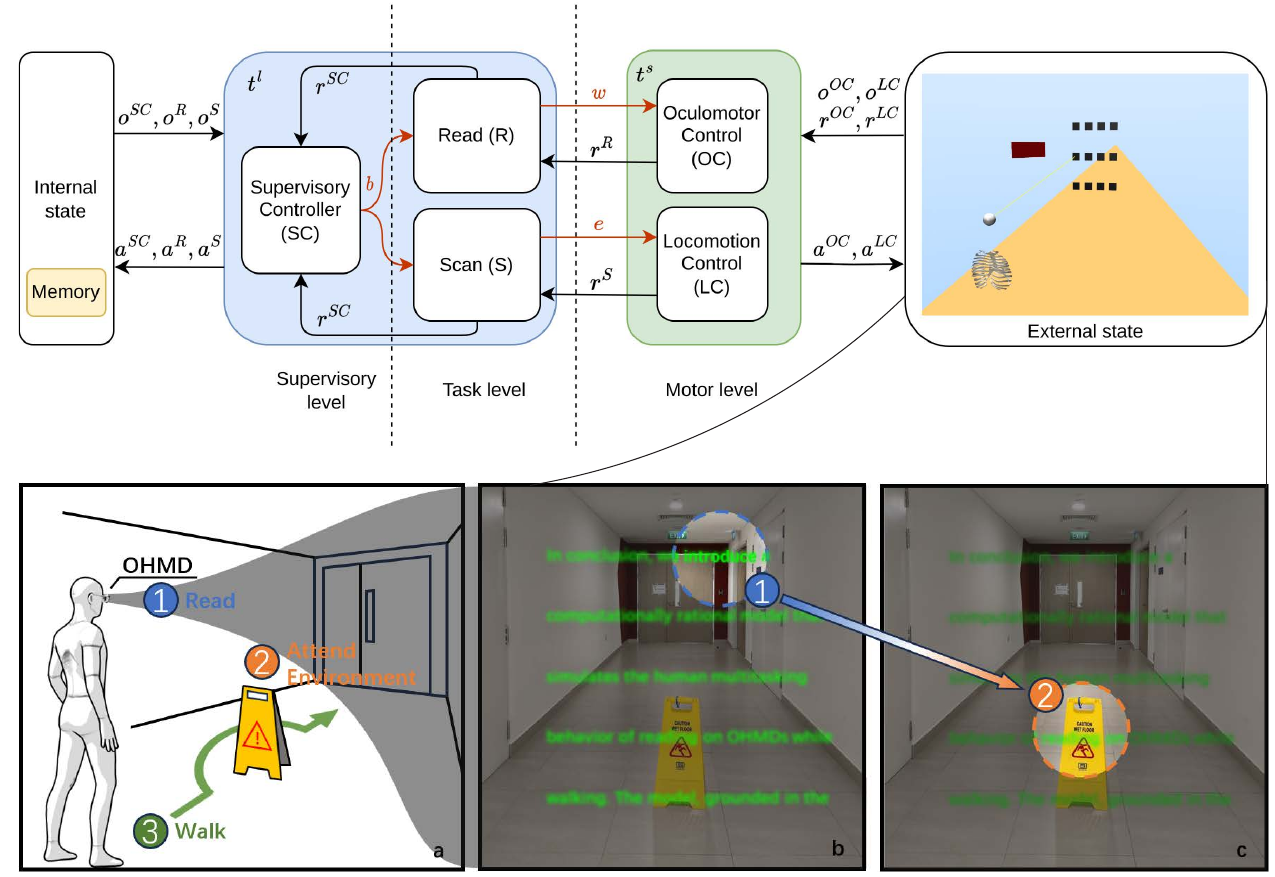}
  \captionsetup{width=1\textwidth}
  \caption{
  \textbf{The Heads-Up Multitasker: a hierarchical, resource-rational model of attention allocation during reading while walking.}
  The model enables a simulated agent to coordinate reading on optical head-mounted displays (OHMDs) with safe locomotion by allocating visual attention under competing task demands. The top panel illustrates the hierarchical reinforcement learning architecture, which decomposes attention control into supervisory, task, and motor levels, allowing interpretable trade-offs between reading and safety to emerge from bounded-optimal control. The bottom panels show the simulated multitasking scenario in the external environment: (a) a third-person view of the agent walking in an environment containing hazards; (b) a first-person view in which the agent reads text presented on OHMDs; and (c) a first-person view in which the agent attends to environmental signage to support safe navigation.
  }
  \Description{}
  \label{fig:hum} 
\end{figure*}

\begin{figure*}[t]
    \centering 
    \includegraphics[width=0.8\textwidth]{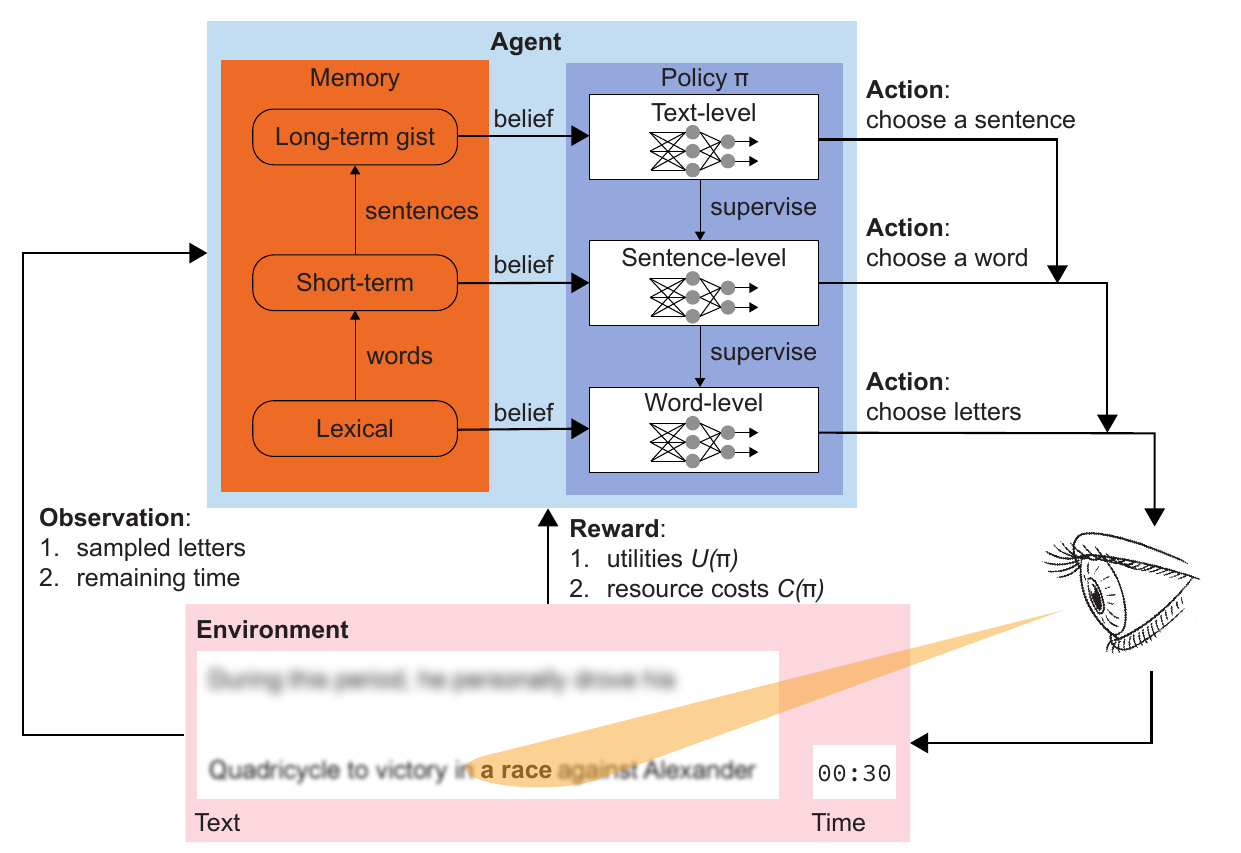} 
    \caption{
    \textbf{Architecture of the resource-rational reading model.}
    The model represents a reader as a sequential decision-making agent that allocates visual attention to maximize text comprehension under constraints of limited memory, visual acuity, and time. Because observations of the text are partial and noisy, the agent maintains probabilistic beliefs at three representational levels: lexical, sentence, and text, supported by corresponding memory systems (lexical store, short-term memory, and long-term gist memory). Decisions are organized hierarchically: a text-level controller guides sentence selection, sentence-level control determines word-level priorities, and word-level control governs eye-movement decisions over visible letters. Rewards combine comprehension utility with resource costs, allowing fixation patterns, skipping, and regressions to emerge as bounded-optimal strategies. This architecture links eye-movement control with memory-driven comprehension, providing an interpretable computational account of human reading behavior relevant to predictive and simulation-based HCI.
    }
    \label{fig:read} 
\end{figure*}


Across my PhD, I have pursued a coherent research trajectory that advances from coarse-grained to fine-grained models of visual attention control. I first studied attention switching in an ecologically grounded multitasking setting—reading on smart glasses while walking—where users must decide when and where to allocate attention between digital content and the physical environment~\cite{bai2024heads}. This scenario captures a ubiquitous and increasingly important everyday challenge as intelligent devices become more seamlessly embedded into daily life, and it highlights the need to balance information intake with safety under real-world constraints.

I then focused on reading as a canonical fine-grained attention task. Reading is a fundamental means of information acquisition, yet eye movements are not a linear sampling process: they are tightly coupled with evolving comprehension and memory states~\cite{rayner1998eye,bairesource}. At this scale, attention control operates at the level of individual letters and words, requiring precise coordination between low-level oculomotor actions and higher-level cognitive processes. I developed a computational model that jointly explains eye-movement control and dynamic, memory-based comprehension within a unified resource-rational principle~\cite{bairesource}. 

Guided by this progression, my research pursues three interrelated objectives:

\begin{itemize}
    \item \textbf{Model multitasking as a resource-rational control process.}  
    I formalize human multitasking as sequential decision-making under uncertainty, where users switch visual attention between smart glasses and the surrounding environment while walking. The agent aims to maximize text comprehension while minimizing safety risks caused by neglecting environmental information, under limited visual-perceptual resources (Figure~\ref{fig:hum}). Key questions include how to specify POMDP components that capture trade-offs between comprehension and safety, and how attentional strategies adapt to contextual factors such as walking speed and interface design (e.g., text spacing and layout).
    \item \textbf{Model text reading as a resource-rational control problem.}  
    I formalize reading as a fine-grained control process in which readers decide where to fixate at the level of letters and words, aiming to maximize comprehension while minimizing oculomotor and time costs under constraints of limited visual acuity, memory capacity, and time (Figure~\ref{fig:read}). Key questions include how POMDP formulations capture adaptations of eye-movement behavior to lexical properties and time pressure, and how these adaptations shape comprehension outcomes.
    \item \textbf{Model individual differences and accessibility in attention allocation during reading.}  
    Building on the reading framework, I investigate how attention-control policies vary across individuals. This includes capturing differences between native and non-native readers, as reported in SB-SAT~\cite{ahn2020towards} and MECO~\cite{kuperman2023text,kuperman2025new}, as well as modeling accessibility-related effects such as those observed in readers with dyslexia using the CopCo corpus~\cite{hollenstein2022copenhagen,bjornsdottir2023dyslexia,reich2024reading}. The goal is to explain individual variation as differences in resource constraints and learned control strategies, rather than ad hoc behavioral rules.
\end{itemize}

Across all projects, I employ a consistent methodological approach based on \emph{hierarchical deep reinforcement learning}~\cite{pateria2021hierarchical}. Hierarchical control decomposes complex attention-allocation problems into multiple interacting timescales, making large and cognitively realistic tasks tractable while preserving interpretability. Although the current instantiations focus on reading-centered tasks, the underlying POMDP formulation does not assume textual structure; state representations and reward functions can be redefined for non-textual tasks such as visual search, dashboard monitoring, or UI exploration.

In the multitasking model (Figure~\ref{fig:hum}), attention control is organized into supervisory, task-level, and motor-level controllers: higher-level policies decide when to switch attention between reading and environmental monitoring, while lower-level policies execute eye and walking movements to acquire task-relevant information. This decomposition allows safety-comprehension trade-offs to emerge naturally as bounded-optimal solutions. 

In the reading model (Figure~\ref{fig:read}), hierarchy reflects the structure of linguistic and memory processes~\cite{bairesource}. A text-level controller governs sentence selection, a sentence-level controller prioritizes words based on evolving comprehension, and a word-level controller selects fixation targets over visible letters. Because observations are partial and noisy, the agent maintains probabilistic belief states over lexical, sentence, and text representations, supported by corresponding memory systems. Policies are learned using deep reinforcement learning (PPO), without direct supervision from human gaze data, enabling fixation patterns, skipping, and regressions to emerge from task optimization under cognitive constraints. This unified methodology supports both explanatory modeling and simulation-based evaluation of human attention behavior in interactive systems.

\section{Results and Contributions}

\begin{figure*}[t]
    \centering 
    \includegraphics[width=\textwidth]{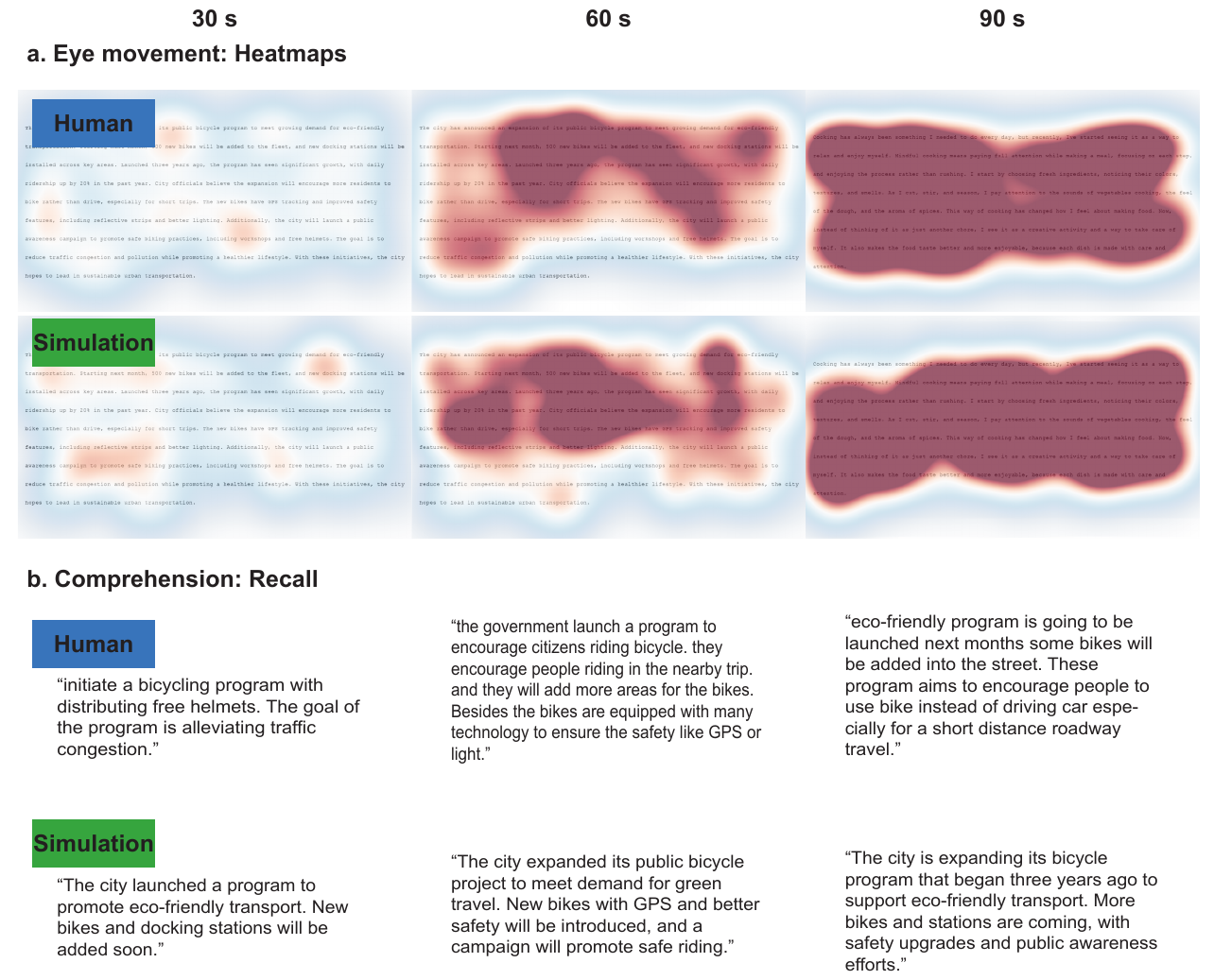} 
    \caption{
    \textbf{Reading behavior under different time constraints (30\,s, 60\,s, 90\,s).} The model and human readers adapt their visual attention strategies as available reading time changes, reflecting resource-rational control of comprehension.
    \textbf{(a) Heatmaps.} With increasing time (from left to right), both humans and the simulation allocate attention more broadly across the text and invest more fixation time, whereas under tight deadlines attention is concentrated on high-coverage regions. 
    \textbf{(b) Recall.} Free-recall responses exhibit the same trade-off, with longer reading time yielding richer gist-level recall and more complete and detailed memory. 
    }
    \label{fig:read models results} 
\end{figure*}

My first line of work introduces the \emph{Heads-Up Multitasker}, the first simulation model to capture how users allocate visual attention while reading on optical head-mounted displays (OHMDs) during walking. Across four empirical and simulation studies, the model reproduces key multitasking phenomena, including systematic attention switching under different walking speeds and task priorities, degraded reading performance due to locomotion-induced perturbations, accurate resumption across OHMD text layouts, and adaptive trade-offs between reading efficiency and walking safety. These results demonstrate that bounded-optimal control provides a principled explanation of how users coordinate perception and action in safety-critical, everyday multitasking scenarios, and they yield predictive insights for the design of safer and more usable OHMD interfaces.

Building on this foundation, my second line of work develops a hierarchical resource-rational model of reading that unifies eye-movement control with memory-driven comprehension. The model explains reading as sequential decision-making that links word-level fixation choices, sentence-level integration, and text-level memory formation within a single computational model. Without training on human eye-tracking data, it reproduces established empirical effects, including systematic variation in fixation durations, skipping, and regressions with lexical difficulty, predictability, and contextual constraints, as well as the influence of text coherence and prior knowledge on comprehension. Extending the model, I collected a new dataset in which participants read under explicit time limits and showed convergent adaptations in both humans and the model: tighter time constraints lead to faster reading, increased skipping, fewer regressions, and reduced comprehension (Figure~\ref{fig:read models results}). This work provides an integrative computational account that jointly explains gaze behavior and evolving memory representations during reading.

Beyond aggregate metrics, the model has also been evaluated at the level of individual scanpath trajectories. Specifically, we compared simulated scanpaths against human participants’ scanpaths on a trial-by-trial basis using Normalized Levenshtein Distance (NLD), a standard metric for scanpath similarity. Importantly, this comparison is conducted between a single simulation rollout and a single human trial, rather than against aggregated human heatmaps. Our resource-rational model achieves the lowest NLD values across time conditions (30~s, 60~s, 90~s), outperforming prior reading models. In addition to trajectory similarity, our approach supports modeling individual differences through its free parameters governing perceptual acuity, memory constraints, and cost sensitivity. By tuning these parameters, the model can reproduce distributions of all metrics, including reading speed, skipping probability, and regression probability that match the observed range of variability across human participants. This enables simulation not only of average reading behavior, but also of specific reader profiles by fitting parameter configurations to individual data. 

While simpler heuristic or rule-based models may capture isolated effects, the resource-rational framework provides a unified generative account that links perception, memory, and task objectives within a single decision-theoretic structure. This enables systematic manipulation of task constraints and design parameters in ways that purely descriptive models cannot readily support.

Together, these projects establish visual attention as a form of hierarchical control, spanning decisions about \emph{what} to attend to across tasks and environments and \emph{how} to sample information at fine perceptual scales. They demonstrate how computational rationality and hierarchical reinforcement learning can produce interpretable and predictive models of complex human behavior without relying on large-scale behavioral imitation. Importantly, this structure enables designers to simulate how attention policies would adapt under alternative interface layouts, time constraints, or resource assumptions before conducting costly empirical studies~\cite{murray2022simulation}. Practically, both research lines deliver open-source simulation environments and datasets that support systematic ``what-if" analyses and scalable evaluation of interaction designs. Collectively, this dissertation positions simulation-based, resource-rational modeling as a foundation for theory-driven and data-efficient interface design in HCI.

\section{Future Directions}

My research will extend resource-rational modeling of visual attention toward increasingly adaptive and interactive settings. From a theoretical perspective, a key next step is to generalize computationally rational control from single-agent behavior to multi-agent and socially situated scenarios. Many real-world tasks, such as collaborative reading in classrooms or shared attention in AR/VR environments require coordination among multiple agents with overlapping but distinct goals. Modeling such settings will enable the study of how attention is allocated not only within individuals but also across groups, opening new avenues for understanding joint cognition and collaborative human–computer interaction.

On the application side, I plan to leverage my models to inform adaptive interface design across a broader range of HCI scenarios. While reading provides a controlled and cognitively rich testbed, the same simulation framework can be extended to interactions with visualizations, images, and complex user interfaces, where users must strategically allocate attention under time and resource constraints. By simulating how users adjust attention during multitasking, time pressure, or high cognitive load, these models can support systematic analysis of design alternatives prior to extensive user studies, and guide the development of time-aware reading interfaces, anticipatory safety prompts in AR systems, and adaptive visual and educational interfaces that respond to predicted user states~\cite{murray2022simulation,bai2024heads}.

Technically, I aim to integrate deep generative models into the simulation pipeline to enhance scalability and realism. Large language models can provide rich contextual predictions to support semantic inference and memory construction~\cite{shi2025chartist}, while diffusion- or transformer-based generative models can be used to synthesize realistic eye-movement patterns or controlled interaction corpora~\cite{bolliger2023scandl,Bolliger2025ScanDL2,deng2023eyettention}. Incorporating these components will allow simulation-based analyses of nuanced comprehension and interaction processes and support interfaces that adapt dynamically to inferred user states.

Finally, within my dissertation roadmap, the remaining milestones focus on extending the reading model to capture individual differences and diverse reader populations. This includes modeling variation across native and non-native readers as well as accessibility-related differences, with planned submissions to UIST and CHI. Together, these directions aim to consolidate a general, extensible framework for modeling and designing attention-aware interactive systems.

At this stage of my dissertation, I am particularly seeking guidance on (i) how to best position resource-rational simulation as a contribution to HCI design practice rather than only cognitive theory; (ii) how much mechanistic detail is necessary for interpretability versus usability by designers; and (iii) how to scope individual-differences modeling within a single dissertation. These questions motivate my participation in the Dissertation Research Roundtable.

\bibliographystyle{ACM-Reference-Format}
\bibliography{reference}










\end{document}